\documentclass[doublecol]{epl2} 

\usepackage{graphicx}
\usepackage{amsmath, amssymb}
\usepackage{lmodern}
\usepackage[T1]{fontenc}
\usepackage{subcaption}
\usepackage[utf8]{inputenc}
\usepackage[english]{babel}
\usepackage{float}
\usepackage{color}
\usepackage{braket}

\title{Uneven rock-paper-scissors models: patterns and coexistence}

\author{J. Menezes \inst{1,2} \and B. Moura \inst{1}  \and T. A. Pereira\inst{3}}

\institute{   
  \inst{1} Escola de Ci\^encias e Tecnologia, Universidade Federal do Rio Grande do Norte\\
Caixa Postal 1524, 59072-970, Natal, RN, Brazil \\                 
  \inst{2} Institute for Biodiversity and Ecosystem
Dynamics, University of Amsterdam, Science Park 904, 1098 XH
Amsterdam, The Netherlands \\
  \inst{3} Departamento de F\'isica Te\'orica e Experimental, Universidade Federal do Rio Grande do Norte, 59078-970, Natal, RN, Brazil
}

\abstract{We study a class of the stochastic May-Leonard models, with three species \mbox{dominating} each other in a cyclic nonhierarchical way, according to the rock-paper-scissors game.  We introduce an unevenness in the system, by considering that one of the species is weaker because of a lower selection probability. The simulation results show that the pattern formation is drastically affected by the presence of the weaker species, with no spiral waves arising immediately from random initial conditions. 
Instead, single-species spatial domains cyclically dominate the entire territory until a region occupied by the weaker species is sufficiently narrow to be crossed by individuals without being selected. This leads to the appearance of spatial patterns responsible for the species coexistence.  We verify that the asymmetry in the selection probabilities leads to different spatial autocorrelation function and average relative species abundances. Finally, we investigate the coexistence probability and show that the surviving species depends on the level of unevenness of the model and the mobility of individuals.}

\begin{document}

\maketitle

\section{Introduction}

\begin{figure}[t]
\centering
\includegraphics[width=50mm]{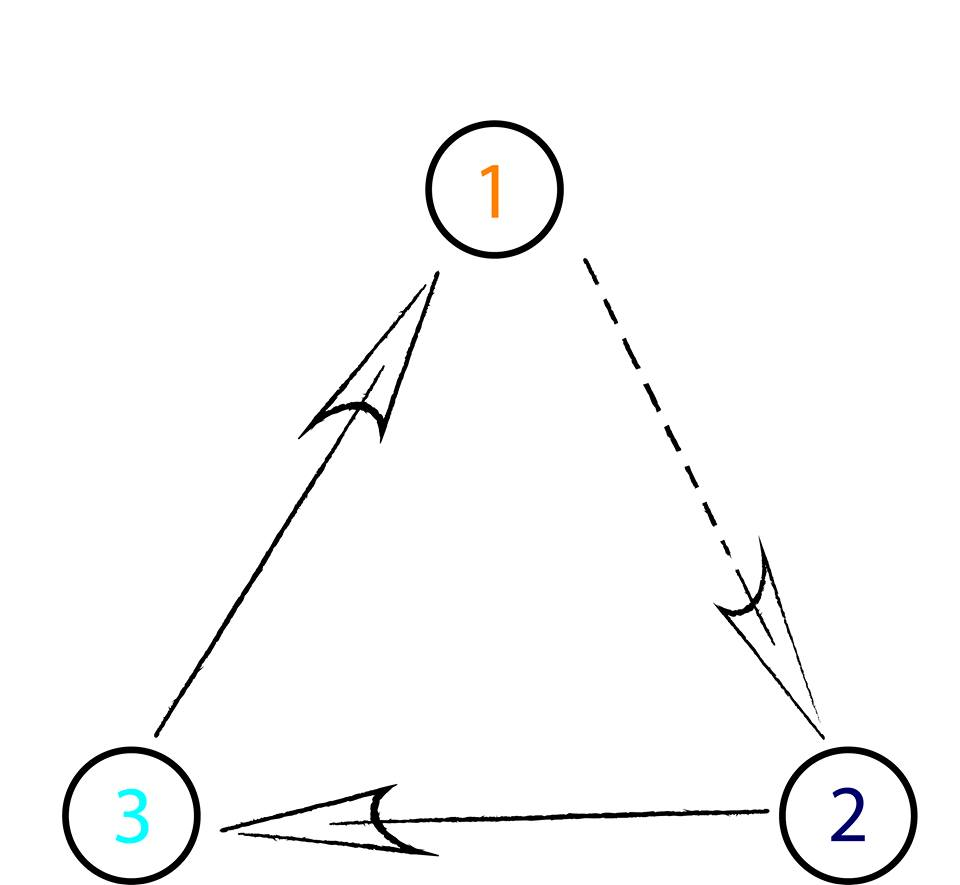}
\caption{Selection interactions among species in an uneven rock-paper-scissors model. The dashed arrow means that species $1$ is weaker than species $2$ and $3$.}
	\label{fig1}
\end{figure}

It is well known that nonhierarchical interactions among species plays a vital role in ecosystem dynamics \cite{Evolutionary}. Furthermore, competition for space allows the formation of spatial patterns that are responsible for the rich biodiversity found in nature (see, for example,  strains of colicinogenic Escherichia coli \cite{Kerr}). To study a three-species system, where species cyclically dominate each other, many authors have used the rock-paper-scissors game. In this model, the spatial interactions are classified as mobility, reproduction, and selection  -  scissors cut paper, paper wraps rock, rock crushes scissors (see Fig. \ref{fig1}). This simple model has been proved to be a powerful tool for the understanding of the ecological processes of biological systems of three cyclically dominant species \cite{Wang,Pereira,Menezes,Losano,Dobramysl,Park,Avelino,Bazeia,Oliveira,Szolnoki,Kei,Frey,Szabo,Boerlijst,Xu}. 

The rock paper scissors model have been largely investigated in the ideal scenario, where all species have an equal probability of interacting. Although this is not necessarily the case for many biological systems (see for example, \cite{Van, Kevin, Berlow, Mila}), numerical results about the role played by species whose strength is somehow reduced, are scarce. 
Some authors have studied this issue in models with a conserved number of individuals, the so-called Lotka-Volterra models \cite{Lotka,Volterra}, 
and others in models without conservation of the total number of individuals \cite{Jang,Dana}, i.e., in models where individuals select each other for gaining local natural resources \cite{Frean,Berr, Park2, Jeppe, Venkat, Quian}.  Specifically, in the May-Leonard type simulations (that have been applied to model real biological systems \cite{Mobilia}),  selection activity allows individuals to control the territory by defeating competitors \cite{May-Leonard,Maynard}.
  
In this letter, we aim to investigate an uneven rock-paper-scissors model where one out of the species is weaker because of a lower probability of selecting (for example, the probability of rock crushing scissors is lower than the probability of scissors or paper acting). Using two-dimensional spatial stochastic simulations, we aim to understand the effects of this unevenness in the spatial pattern formation and the population dynamics. Lastly, we will focus on the coexistence probability in terms of the strength ratio between the weaker species and the others.
\section{The model}

We consider a cyclic rock-paper-scissors model, that is a class of the May-Leonard models\cite{May-Leonard}. In our simulations, individuals may select one of their eight immediate neighbors, switch spatial positions, and produce offspring if there is empty space in their vicinity.  In summary, the possible interactions 
are selection ($ i\ j \to i\ \otimes\,$), mobility ($ i\ \odot \to \odot\ i\,$), and reproduction ($ i\ \otimes \to ii\,$), with $i \neq j = 1, 2, 3$.
While ($\otimes$) means an empty space, $\odot$ may be either an individual of any species or an empty space. 

\begin{figure*}
\centering
    \begin{subfigure}{.2\textwidth}
        \centering
        \includegraphics[width=33mm]{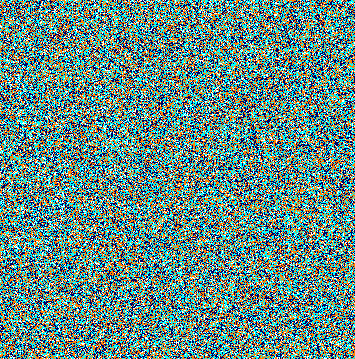}
        \caption{}\label{fig2a}
    \end{subfigure} %
       \begin{subfigure}{.2\textwidth}
        \centering
        \includegraphics[width=33mm]{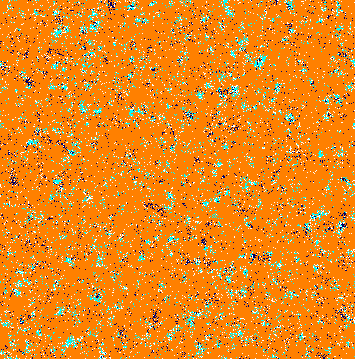}
        \caption{}\label{fig2b}
    \end{subfigure} %
   \begin{subfigure}{.2\textwidth}
        \centering
        \includegraphics[width=33mm]{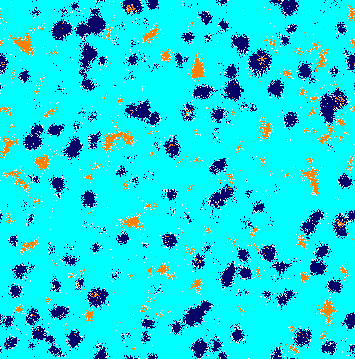}
        \caption{}\label{fig2c}
    \end{subfigure} %
   \begin{subfigure}{.2\textwidth}
        \centering
        \includegraphics[width=33mm]{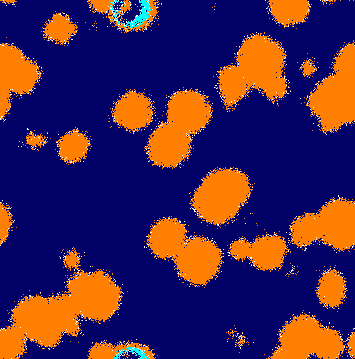}
        \caption{}\label{fig2d}
    \end{subfigure} %
   \begin{subfigure}{0.2\textwidth}
        \centering
        \includegraphics[width=33mm]{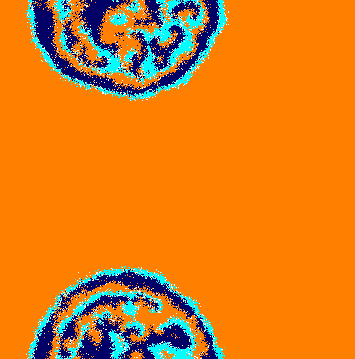}
        \caption{}\label{fig2e}
    \end{subfigure} %
   \begin{subfigure}{.2\textwidth}
        \centering
        \includegraphics[width=33mm]{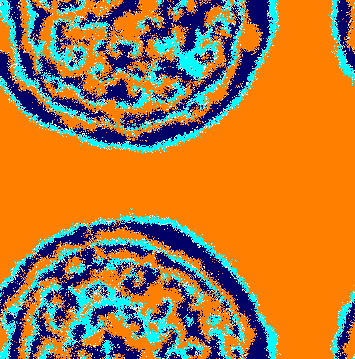}
        \caption{}\label{fig2f}
    \end{subfigure} %
    \begin{subfigure}{.2\textwidth}
        \centering
        \includegraphics[width=33mm]{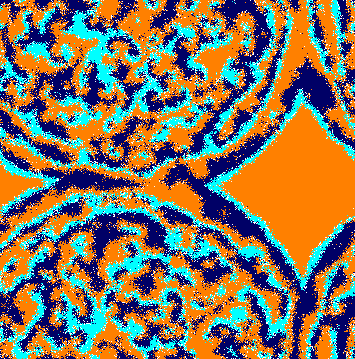}
        \caption{}\label{fig2g}
    \end{subfigure} %
    \begin{subfigure}{.2\textwidth}
        \centering
        \includegraphics[width=33mm]{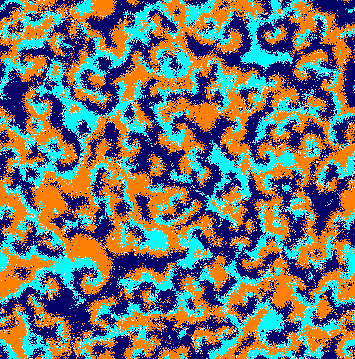}
        \caption{}\label{fig2h}
    \end{subfigure}
\caption{Spatial patterns of a $750^2$ simulation for $m=\,r\,=\,c\,=\,1/3$ and $\kappa_1 = 0.5$. Figures a, b, c, d, e, f, g, h, show the initial conditions and snapshots captured after
$28$, $68$, $186$, $502$, $702$, $902$, and $2000$ generations, respectively. Empty spaces and individuals of species $1$, $2$, and $3$, are represented by white, orange, dark blue, and cyan dots, respectively.}
  \label{fig2}
\end{figure*}


The simulations were performed in square lattices of  $\mathcal{N}$ sites, with boundary periodic conditions. We assumed random initial conditions, where each grid site is empty or contains at most one individual of an aleatory species.
Initially, the total numbers of individuals of every species ($I_i$, for $i=1,2,3$) are the same. 
At each timestep, the simulation algorithm follows three steps: i.) selecting a random occupied grid point to be the active position; ii.) drawing one of its eight neighbour sites to be the passive position; iii.) randomly choosing an interaction to be executed by the individual at the active position.  If the content of the active and passive positions does not match the raffled interaction, the code repeats the three steps. 
This implies that, once the active and passive positions have been sorted out, the probability of implementing an interaction is less than $1$:  $m+\kappa\,c<1$, if individuals of species $i$  and $i+1$ occupy the active and passive positions, respectively; $m+r<1$, if an individual of any species $i$  is present in the active position and the passive position is empty; $m<1$, if individuals of species $i$  and $i-1$ occupy the active and passive positions, respectively;  $m<1$, if individuals of the same species occupy the active and passive positions.  Our time unit is called generation, that is the necessary time to $\mathcal{N}$ interactions to occur.

Different from previous works (see, for example, \cite{Mobilia,Mobilia2}), where mobility probability is defined in terms of the total number of grid points, we consider that $m$ does not depend on the lattice size.  In other words, irrespectively of $\mathcal{N}$, the average area (number of grid points) explored by one individual is the same. Therefore, changing $\mathcal{N}$ for the same set of parameters means assuming different system sizes.
\begin{figure}
\centering
      \begin{subfigure}{0.2\textwidth}
        \centering
        \includegraphics[width=33mm]{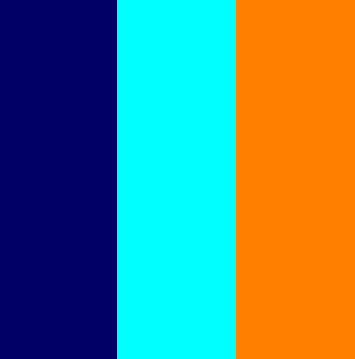}
        \caption{}\label{fig4a}
    \end{subfigure} %
   \begin{subfigure}{.2\textwidth}
        \centering
        \includegraphics[width=33mm]{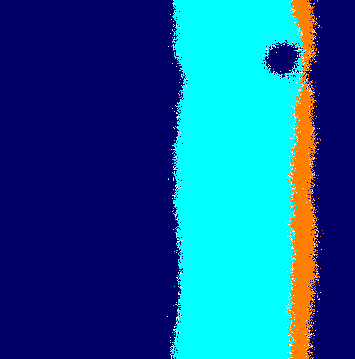}
        \caption{}\label{fig4b}
    \end{subfigure} %
    \begin{subfigure}{.2\textwidth}
        \centering
        \includegraphics[width=33mm]{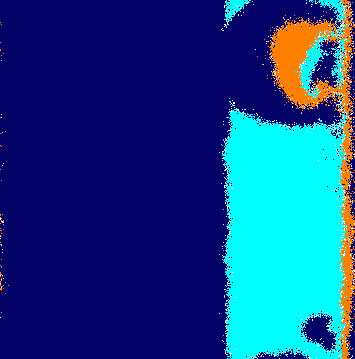}
        \caption{}\label{fig4c}
    \end{subfigure} %
    \begin{subfigure}{.2\textwidth}
        \centering
        \includegraphics[width=33mm]{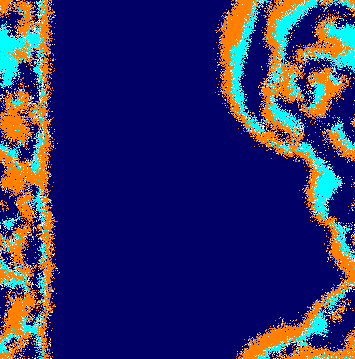}
        \caption{}\label{fig4d}
    \end{subfigure}
\caption{Snapshots obtained from a $600^2$ simulation with $m\,=\,r\,=\,c\,=\,1/3$, and $\kappa_1 = 0.5$. Fig. \ref{fig4a} shows how the species were initially distributed, while Fig. \ref{fig4b}, Fig. \ref{fig4c}, Fig. \ref{fig4d} depict the spatial patterns after $675$, $750$, and $825$ generations.}
\end{figure}

\section{Results}
\subsection{Spatial patterns}
We performed a large number of simulations to understand how a weaker species interferes in the dynamics of the rock-paper-scissors model.  Figure \ref{fig2} shows snapshots of the spatial patterns captured from a $750^2$ lattice, with a timespan of $2000$ generations. 
Orange, dark blue, and cyan dots represent individuals of species $1$, $2$, and $3$, respectively, whereas white dots show the empty sites. From left to right, and from top to bottom, the panels show the initial conditions and the spatial patterns after $26$, $66$, $184$, $500$, $700$, $950$, and $2000$ generations, respectively. The temporal evolution of the spatial patterns is shown in video in \cite{video1}.

Because of the random initial distribution (Fig. \ref{fig2a}),
there is a high selection activity in the beginning of the simulation.  Thereat, individuals of a single species form groups working together to expand the area that they occupy. Due to the selection interaction unevenness among the species, there is no spiral pattern formation at this first stage of the simulation. Instead,  the weaker species dominates almost all territory. This happens because individuals of species $1$ select less, allowing the population of species $2$ to grow more than others, limiting the population growth of species $3$. Therefore, there are fewer individuals to select the species $1$, which spreads  (Fig. \ref{fig2b}). 
The high abundance of species $1$ gives an opportunity to the few remaining individuals of species $3$ to reproduce, forming expanding spatial regions (Fig. \ref{fig2c}).  
Then, the same happens with species $2$, which reproduces rapidly due to the plenty of individuals of species $3$ (Fig. \ref{fig2d}). 
The continuous creation of expanding single-species domains can be interrupted by the formation of local spiral patterns. This process results from stochastic movements of individuals of species $2$, which cross the area dominated by individuals of species $1$ without being caught. This generates spiral waves that spread  throughout the lattice (Figs. \ref{fig2e},  \ref{fig2f}, \ref{fig2g}, and \ref{fig2h}).

To see this effect in more details, we run a $600^2$ lattice simulation starting from the initial conditions depicted by Fig. \ref{fig4a}, where each species occupies one-third of the grid. Because of the periodic boundary conditions, the individuals are initially confined in single-species rings. As soon as the simulation starts, the cyclic selection interaction leads to the movement of the rings around the toric surface from left to right.  Figure \ref{fig4b} shows that the selection unevenness causes the narrowing of the ring occupied by species $1$, which attacks less than it is attacked.
To calculate how the average ring widths change in time, 
we consider the area of the ring occupied by species $i$, which is defined as $I_i$, the total number of individuals of species $i$.   Taking into account that the torus cross section perimeter $\sqrt{\mathcal{N}}$ does not change in time, the temporal changes in the ring width, $\delta_i$, are given by 
\begin{equation}
\dot{\delta_i}\,= \,\mathcal{N}^{-1}\,\dot{I_i},
\end{equation}
where the dot stands for the time derivative.  Computing the temporal variation of the total number of individuals of each species, we concluded that $\dot{\delta_1}=-0.29$. Besides, the ring of species $2$ enlarges with the same rate, that is, $\dot{\delta_2}=0.29$, while the ring occupied by species $3$ has a constant average width.

In addition, stochastic fluctuations occurring on the domain's borders generate short ways that allow individuals of species $2$ to cross the entire ring of species $1$ without being selected (this is more likely for smaller $\kappa_1$). Once they reach the spatial domains of species $3$, they start proliferating in expanding clusters (Fig. \ref{fig4b}). 
Further, the abundance of individuals of species $2$ permits a movement of groups of individuals of species $1$ followed by individuals of species $3$ in the opposite direction to ring rotation.
This leads to the appearance of spiral patterns (Fig. \ref{fig4c}) that dominate the entire lattice (Fig. \ref{fig4d}). 
Video \cite{video2} shows another realisation of a $900^2$ simulation starting with the initial conditions depicted in Fig. \ref{fig4a} . 

If the initial conditions are not random, spiral waves only appear when groups of the three species are in contact. The videos in \cite{video3,video4,video5} show some examples of $300^2$ simulations running until $800$ generations, for $r = c = m = 1/3$, and $k_1 = 0.5$.  In the video in \cite{video3}, square domains of species 1, 2 and 3  are initially in contact, leading to the immediate arising of spiral waves. In the video in \cite{video4},  circular domains of each species are assumed. Although the species are not initially in contact, the narrowing of the domain of species $1$ allows individuals of species $2$ to reach the circular domain of species $3$, leading to the formation of spirals (this process is similar to Fig. 3). Finally, in the video in  \cite{video5}, circular domains are again assumed in the initial conditions, but the domain thickness is larger than in video in \cite{video4}. As a consequence, the narrowing of the domain occupied by species $1$ is not fast enough to allow the process seen in video in \cite{video4}. In this case, spiral waves do not appear, causing the extinction of species $1$ and $3$.  As one sees in the videos in \cite{video2}, \cite{video3} and \cite{video4}, the necessary time to spiral waves appear depends on the initial configuration. This is the time needed for groups of three species to get in contact, giving rise a local uneven rock-paper-scissors dynamics.

\subsection{Relative Species Abundance}
\begin{figure}
\centering
\includegraphics[width=80mm]{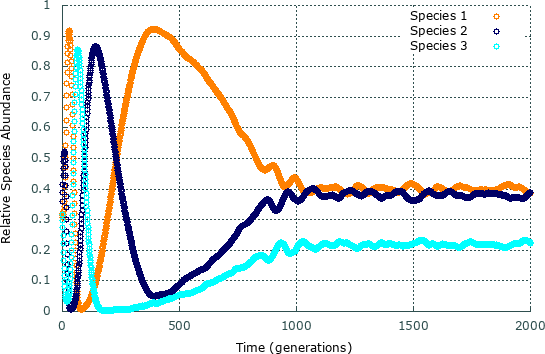}
\caption{Temporal changes in the relative species abundance during the simulation presented in Fig. 2. Orange, dark blue and cyan lines represent the species $1$, $2$, and $3$, respectively.}
	\label{fig3}
\end{figure}

To investigate the population dynamics, we calculated the relative species abundance as the fraction of individuals of species $i$, i.e., 
\begin{equation}
\rho_i\, =\, \frac{I_i}{\sum_{j=1}^{3}{I_j}}.
\end{equation}
Figure  \ref{fig3} depicts $\rho_i$ for the timespan of the simulation presented in Fig. \ref{fig2}.

In the first period of the simulation, the oscillations on species populations are accentuated, revealing the arising and growth of spatial areas inhabited by single species shown in Fig. 2.  Hereafter, the spiral pattern appearance drives the system to a rock-paper-scissors dynamics, where the relative species abundances $\rho_i$ oscillate around an average value over time, $\bar{\rho_i}$. After $1000$ generations, it is possible to see that the population size of species $1$ (the weaker species) and $2$ are very close, while the population size of species $3$ is much inferior. This happens because i) individuals of species 1 select little, allowing the population of species $2$ to grow; ii) plenty of individuals of species $2$ chase individuals of species $3$, which means that the population of species $3$ is controlled at low density ; iii) due to the low population of species $3$, the number of attacks on species $1$ is reduced, which maintains the population of species $1$ with high values.  In summary, for $\kappa_1 = 0.5$,  
the average relative species abundances are: $\bar{\rho_1}\, \approx 0.40$, $\bar{\rho_2}\,\approx 0.38$, and $\bar{\rho_3}\,\approx\,0.22$.

\subsection{Autocorrelation Function}

\begin{figure}
\centering
\includegraphics[width=80mm]{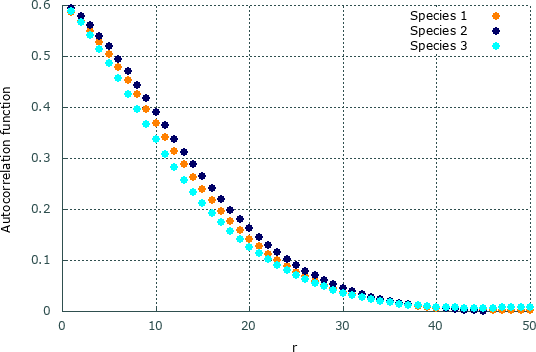}
\caption{Spatial autocorrelation function for species $1$ (orange), $2$ (dark blue), and $3$ (cyan). The results were obtained by using the spatial distribution of individuals in the snapshot depicted by Fig. \ref{fig2h}.}
	\label{fig5}
\end{figure}

We now aim to quantify the effect of the selection unevenness on spatial distribution of species by means of the spatial autocorrelation function
\begin{equation}
C(r') = \sum_{|\vec{r}'|=x+y} \frac{C(\vec{r}')}{min (2N-(x+y+1), (x+y+1) }.
\end{equation}
This function is computed from the Fourier transform of the spectral density as 
\begin{equation}
C(\vec{r}') = \frac{\mathcal{F}^{-1}\{S(\vec{k})\}}{C(0)},
\end{equation}
where the spectral density $S(\vec{k})$ is given by
\begin{equation}
S(\vec{k}) = \sum_{k_x, k_y}\,\varphi(\vec{\kappa}),
\end{equation}
with $\varphi(\vec{\kappa}) = \mathcal{F}\,\{\phi(\vec{r})-\braket{\phi}\}$. The function $\phi(\vec{r})$ represents the species in the position $\vec{r}$ in the lattice (we assumed $0$, $1$, $2$, and $3$, for empty sites, and individuals of species $1$, $2$, and $3$, respectively). 

Figure \ref{fig5} depicts the autocorrelation function for species $1$, $2$, and $3$ for the snapshot showed in Fig. \ref{fig2h}. 
Accordingly, the characteristic length $\ell$, which we defined as $C( r' = \ell) = 0.15$, 
is different for each species: $\ell_1 = 20.47$, $\ell_2 = 21.70$, and $\ell_3 = 19.44$. This result shows a asymmetric formation of spirals, with a larger spatial correlation  
between individuals of species $2$.

\subsection{Coexistence Probability}

The spatial pattern formation showed in Fig. \ref{fig2h} is 
responsible for the species coexistence. The probability of occurrence of a stochastic event that generates spiral patterns is proportional to the number of grid points. This means that the larger the lattice is, the more probably the species coexist. 
Now, we aim to study how the coexistence probability changes (for a fixed grid size), when we vary the weaker species strength and mobility of individuals. 
To this purpose, we run one million $50^2$ simulations until $250$ generations.  

\begin{figure}
\centering
\includegraphics[width=75mm]{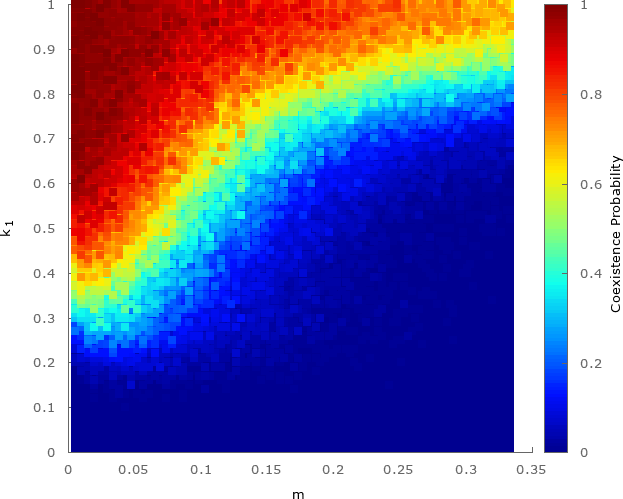}
\caption{Coexistence probability as a function of the weaker species strength factor and the mobility parameter. The colour bar indicates the probability for each pair of parameters. Each dot was obtained by averaging $100$ simulations for $r=c=(1-m)/2$.}
	\label{fig6}
\end{figure}

Figure \ref{fig6} shows the coexistence probability for $0\,<\kappa_1\,\leq 1$, and $0\,<\,m\,\leq1/3$.
Each dot shows the probability averaged from a set of $100$ simulations with different initial conditions.  The results showed that, as the weaker species strength factor approaches to zero, the chances of the coexistence decreases. This happens because for smaller $\kappa_1$ the cyclical equilibrium is increasingly compromised. 
Besides, the coexistence is dependent on the mobility probability:
for large $m$, individuals can go further on the lattice, increasing the outcome of the presence of a weaker species in the system.  

To quantify the transition between coexistence and extinction regimes, we defined the critical strength factor $\kappa_c$ as $\kappa_1$ whose coexistence probability is $50\%$. 
According to the results obtained from Fig. \ref{fig6},  the best fit for the critical strength factor is given by (the errors are in parenthesis)
\begin{equation}
\kappa_c\, =  0.68 \, (\pm 0.006)\, \tanh {(2\,\pi m)}\, +\, 0.23\, (\pm 0.004),
\end{equation}
with $m\,\leq\,1/3$. For $\kappa_1 < \kappa_c$, the disappearance of two species becomes more likely. 

\begin{figure}
\centering
\includegraphics[width=75mm]{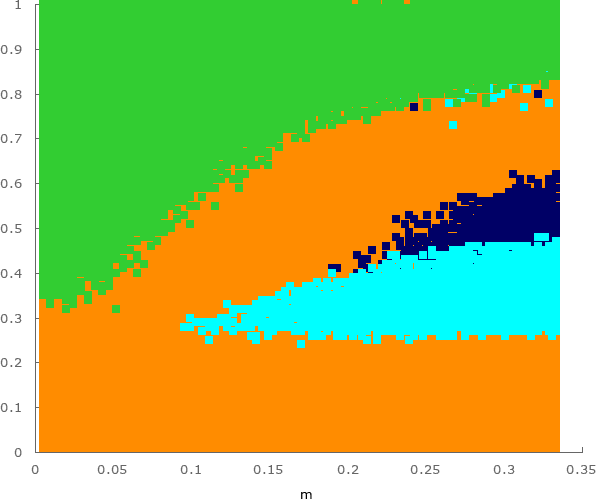}
\caption{The most frequent surviving species as a function of the weaker species strength factor and the mobility parameter. Orange, dark blue and cyan dots shows the pair of parameters leading to extinction where only individuals of species $1$, $2$, and $3$ stay alive. Green dots represent coexistence. Each dot shows the mode of $100$ simulations with different initial conditions, for $r=c=(1-m)/2$.}
    \label{fig7}
\end{figure}

Figure \ref{fig7} shows the species that survives most often from each set of $100$ simulations shown in Fig.  \ref{fig6} (for each pair of parameters $\kappa_1$ and $m$). Green dots show the cases where coexistence happens more often, while orange, dark blue and cyan indicate the mode of the only survivor. The results presented in this letter are in disagreement with  Ref. \cite{Frean}.  The reason is that the authors considered a different model, with a conservation law for the total number of individuals on the lattice, so-called Lotka-Volterra model [16, 17]).  In that case, there are no empty spaces on the lattice because predation and reproduction interactions happen simultaneously.

\section{Conclusions}

We simulate a class of May-Leonard models with three species following the rock-paper-scissors dynamics. Considering that one of the species has difficulties in selecting individuals, the cyclic stability might be compromised. The unevenness on the species strength is reflected on the pattern formation: spiral waves do not arise immediately from the random initial conditions but later, taking the territory occupied by large single-species domains. Utilising the spatial autocorrelation function of each species, and the temporal changes on the relative species abundance in the system, we verified an asymmetry in the way the species behave in space as well as the population dynamics.

Despite the presence of a weaker species in the system, coexistence is not destroyed for large lattices. However, if there is no room for the spiral pattern to form, the species may be in risk of extinction.  This happens because mobility probability is independent of the total number of grid points, which means that, increasing $N$ creates larger physical systems with spatially uncorrelated regions.  Therefore, for larger $\mathcal{N}$,  there is more space for groups of the three species to get in contact, giving rising spiral waves.

Running a huge number of simulations for $m \leq 1/3$, we noticed that, for small mobility probability, coexistence is expected even if the weaker species has half the strength of the others. On the other hand, irrespectively of the mobility probability, if the weaker species strength factor is less than one-third, the biodiversity is destroyed.  Lastly, when extinction is more likely, the weaker species is not necessarily the only surviving species. 

The knowledge of the conditions where the uneven tritrophic interaction maintains the coexistence may help ecologists to understand how the model can be extrapolated to explain real scenarios where players are not able to play fairly.  For example, a seasonal variation on the capacity of one species to compete by the natural resources can lead to the end of the system, or even to an adjustment on the population dynamics.

\acknowledgments
We thank CNPq, ECT, Fapern,  the Netherlands Organisation for Scientific Research (NWO) for financial and computational support.  JM acknowledges support from NWO Visitor's Travel Grant 040.11.643.

\end{document}